\documentclass[aps,pre,floats,showpacs]{revtex4}

\usepackage{epsfig}
\usepackage{color}
\usepackage{bm}
\usepackage{latexsym}
\usepackage{times}

\begin{document}
\newcommand{\hide}[1]{}
\newcommand{\tbox}[1]{\mbox{\tiny #1}}
\newcommand{\half}{\mbox{\small $\frac{1}{2}$}}
\newcommand{\sinc}{\mbox{sinc}}
\newcommand{\const}{\mbox{const}}
\newcommand{\trc}{\mbox{trace}}
\newcommand{\intt}{\int\!\!\!\!\int }
\newcommand{\ointt}{\int\!\!\!\!\int\!\!\!\!\!\circ\ }
\newcommand{\eexp}{\mbox{e}^}
\newcommand{\bra}{\left\langle}
\newcommand{\ket}{\right\rangle}
\newcommand{\EPS} {\mbox{\LARGE $\epsilon$}}
\newcommand{\ar}{\mathsf r}
\newcommand{\im}{\mbox{Im}}
\newcommand{\re}{\mbox{Re}}
\newcommand{\bmsf}[1]{\bm{\mathsf{#1}}}
\newcommand{\mpg}[2][1.0\hsize]{\begin{minipage}[b]{#1}{#2}\end{minipage}}

\title{Cognitive hierarchy theory and two-person games}

\author{Carlos Gracia-L\'azaro $^{1}$, Luis M. Flor\'{\i}a $^{1,2}$ and Yamir Moreno $^{1,3,4}$}

\affiliation{
$^1$Institute for Biocomputation and Physics of Complex Systems (BIFI), University 
of Zaragoza, Zaragoza 50009, Spain \\
$^2$Department of Condensed Matter Physics, University of Zaragoza, Zaragoza 50009,
Spain \\
$^3$Department of Theoretical Physics, University of Zaragoza, Zaragoza 50009,
Spain \\
$^4$Complex Networks and Systems Lagrange Lab, Institute for Scientific 
Interchange, Turin, Italy
}

\date{\today}

\begin{abstract}
The outcome of many social and economic interactions, such as stock-market transactions, is strongly determined by the predictions that agents make about the behavior of other individuals. Cognitive Hierarchy Theory provides a framework to model the consequences of forecasting accuracy that has proven to fit data from certain types of game theory experiments, such as Keynesian Beauty Contests and Entry Games. Here, we focus on symmetric two-players-two-actions games and establish an algorithm to find the players' strategies according to the Cognitive Hierarchy Approach. We show that the Snowdrift Game exhibits a pattern of behavior whose complexity grows as the cognitive levels of players increases. In addition to finding the solutions up to the third cognitive level, we demonstrate, in this theoretical frame, two new properties of snowdrift games: i) any snowdrift game can be characterized by only a parameter -- its class, ii) they are anti-symmetric with respect to the diagonal of the pay-off's space. Finally, we propose a model based on an evolutionary dynamics that captures the main features of the Cognitive Hierarchy Theory.
\end{abstract}

\pacs{
}

\maketitle

\section{Introduction}

Many real-life situations in human societies imply interactions in which the
results of one person's choices depend not only on his own behavior but also on the choices of the other
individuals involved. In these situations, it is usually assumed that people 
behave strategically,
taking into account the likely responses of the other participants who might have an impact on their own
benefit. Most theories of behavior assume rationality; perfect rationality is based on
two assumptions, namely that agents form correct beliefs about other agents' behavior and 
that they choose those actions that 
maximize their own utility functions. Otherwise, when the rationality of agents is limited by practical elements
such as cognitive and time limitations or the tractability
of the decision problem, it is said that
there is bounded rationality. Bounded rationality does not involve a maximization
of the outcome, since agents can make wrong assumptions about the behavior of others. Indeed,
the level of accuracy in the predictions on the other agents' actions plays a key role in some situations
such as stock-market transactions. These kinds of situations in which agents' outcome is strongly
determined by their predictions are captured in the Keynesian Beauty Contests \cite{Keines1936} and Entry Games.  In the p-Beauty Contest Game \cite{Moulin1986}, participants have to simultaneously pick a number between 0 and 100. The winner of the game is the person(s) whose chosen number is closest to $p$ times the average of all selections, where $0<p<1$, typically $p=2/3,1/2$. Entry games are anti-coordination
games in which agents have to decide whether or not to incur a cost to enter a 
market \cite{Farrell1987,Darrough1990,Krugman1995,Erev1998,Rapoport2000}. The entrants' profits will be
positive if the other agents do not enter, but otherwise can turn out to be negative. In these
games, when players act overconfidently, that is, assuming that the other players do
not act with such refined reasoning as they do, the players
are not in equilibrium. Cognitive Hierarchy theories capture this 
behavior by classifying the players according to their degree of reasoning in forming expectations of
others
 \cite{Stahl1995,Costa-Gomes2001,Camerer2004,CrawfordAmEcRev2007,CrawfordEconometrica2007,Costa-Gomes2011,Hossain2013}. These
theories are characterized by a distribution
of the number of iterated reasoning steps that players can
do, \textit{i.e.}, the distribution of players' levels. While zero-step (level-$0$) players
just play at random, higher level players assume they are playing against players who do
fewer reasoning steps than they do. The game can be solved by knowing the distribution of
players' levels and the assumptions of players about the distribution of their opponents' levels.
Camerer \textit{et al.} found that a Poisson distribution fits experimental data from
many different games  \cite{Camerer2004}. A Poisson distribution is fully characterized by its mean, in this case the
average number of reasoning steps, and they found that an average of $1.5$ steps fits many experimental
data. This value implies a fast decay: while 81\% of players do, at most, two reasoning steps, only
1\% of them do more than four steps
, which reflects
the limitations of memory and reasoning ability. 

Secondly, socially relevant situations usually involve social 
dilemmas where individuals profit from selfishness 
at the expense of collective welfare \cite{Dawes1980,Kollock1998,VanLange2013},  
as well as coordination and anti-coordination quandaries where 
all parties can maximize their benefits by making mutually
consistent decisions \cite{Smith1982,Cooper1998,Skyrms2003,Sugden2005,Bramoulle2007}. These situations
have been widely studied in different disciplines ranging from economics, sociology, political
science to psychology, by using the framework of Game Theory to understand how people approach conflict
and cooperation under modeling conditions \cite{Myerson1991,Gintis2009,Sigmun2010,Perc2008}. In this sense, experimental research has shown that, when people face these situations in game theory experiments, they do not always exhibit rational behavior, either because they do not try to optimize their benefit exclusively or because 
of individual or practical limitations \cite{Kagel1997,Camerer2003}.

Here, we focus on a set of two-players-two-actions
games that capture two important features
of social interaction, namely, the dilemma between self-interest and the common good
and coordination issues \cite{Rapoport1966}. In line with previous literature, we
refer to these two actions as cooperation, when the choice transcends self-interest and
concentrates on the welfare of collective, or defection, when it is focused on promoting self-interest.
This set of games includes the Stag Hunt (SH) \cite{Skyrms2003}, the Snowdrift Game (SG) \cite{Smith1982,Sugden2005},
and the Prisoner's Dilemma (PD) \cite{Rapoport1966,Axelrod1981}. SH is a coordination game that describes a
conflict between safety and social cooperation,
the greatest benefit for both players is obtained when both
choose to cooperate, but against a defector the best action is to defect,
so that cooperation is the most advantageous and risky choice. SG is an
anti-coordination game where the greater individual benefit is obtained
by defecting against a cooperator, but players are penalized when both choose to defect,
so that it is always more advantageous to choose the opposite action of your opponent.
In PD, a player always get the highest individual benefit by defecting,
while the greater collective benefit is obtained when both cooperate. For completeness, we also 
study the harmony game (HG), where the best choice
is always to cooperate, regardless of the opponent's behavior; therefore,
there are no tensions between individual and collective benefits. An arrangement of these
four games has
been experimentally studied, finding that players can be classified into four basic personality
types: optimistic, pessimistic, trusting and envious, with only a small fraction of undefined
subjects \cite{Poncela2016}. Although
some of these four games, particularly SH, have being solved according to the Cognitive Hierarchy approach \cite{Camerer2004},
and the solutions for PD and HG are straightforward, SG presents an intricate pattern of
behavior as the cognitive level of the players grows. In this study, 
we establish an algorithm to solve the SD case: in addition to analytically solving
it up to the third cognitive level, we show some symmetries valid for all levels.

We round off this study by exploring the situation in which players can change their guesses about how cognitive levels are distributed in the population. Evolutionary Game Theory is concerned with entire populations of agents
that can choose actions according to some strategies in their interactions with
other agents \cite{Hofbauer1998,Hintis2000,Hofbauer2003}. We propose a model based
on an evolutionary dynamics, in which the agents of
a population interact 
among them 
through the above described games. In this iterated model, the agents do not have
any information of the other players, neither regarding their payments nor on their actions,
but only one-step memory of their own
payment. According to this dynamics, the players make attempts to modify their
assumptions about the distribution of the cognitive levels of the other players,
allowing them to change their assumptions in case their payment decreases. We numerically solve the
model using Monte Carlo simulations, finding patterns of behavior compatible with our theoretical predictions.

\section{Results}

\subsection{Preliminary concepts}
\subsubsection{Two-person games}

Symmetric two-players-two-actions games can be expressed by means of its payoff matrix, where rows represent focal player's actions, columns represent opponent's actions, and the corresponding matrix element is the payoff received by the focal player:

\begin{eqnarray}
\bordermatrix{
 & C & D \cr
C & R & S \cr
D & T & P \cr}\;\;.
\label{payoffMatrix} 
\end{eqnarray}

Actions $C$ and $D$ are usually referred to as cooperation and defection respectively. 
Each player chooses one of the two available actions, cooperation or defection. A
cooperator receives $R$ when playing with a cooperator, and $S$
when playing with a defector, while a defector earns $P$ when
playing with a defector, and $T$ (temptation) against a
cooperator. When $T>R>P>S$, the game is a Prisoner's Dilemma (PD), while if $T>R>S>P$
it is called Snowdrift Game (SG), also Chicken or Hawks and Doves. Otherwise, if $S>P$ and $R>T$ the game is referred to as
Harmony Game (HG), while if $R>T>P>S$ it is called Stag Hunt Game (SH). 

We consider a well mixed population of $N$ agents. According to the payoff matrix (\ref{payoffMatrix}), a cooperator will receive a payoff $N_cR+(N-N_c)S$, where $N_c$ is the number of cooperators,
while a defector will receive $N_cT+(N-1-N_c)P$. A given player will obtain a higher payoff by cooperating than defecting whenever
$cR+(1-c)S>cT+(1-c)P$, where $c$ is the fraction of cooperators in the population, excluding himself. That is, there is a threshold $S_{th}$ for
the parameter $S$:

\begin{equation}
S_{th}(T,P,R;c)=\frac{P+c(T-P-R)}{1-c}\;\;,
\label{S_threshold}
\end{equation}

\noindent above which a player will obtain a higher payoff by cooperating than by defecting.

In order to have a two dimensional representation of the parameter space of the four types of game described above, let us fix the values of the payoff parameters
$P=1$, $R=2$. By varying the values  of $T$ and $S$ over the ranges $T \in [1,3]$ and $S \in [0,2]$, the plane $(T,S)$ can
be divided into four quadrants, each one corresponding to a different type of
game: HG ($T<2,\, S>1$), SG ($T>2,\, S>1$), SH ($T<2,\, S<1$) and PD ($T>2,\, S<1$). According to
these values, equation (\ref{S_threshold}) becomes:

\begin{equation}
S_{th}(T;c)=\frac{1+c(T-3)}{1-c}\;\;.
\label{S_threshold_Tc}
\end{equation}

Note that, for fixed $T>2$, $S_{th}$ is an increasing function of $c$, while it is decreasing for $T<2$. This observation will be crucial in some of the arguments below in next subsections.

\begin{figure}[h]
\centering
\includegraphics[width=15cm]{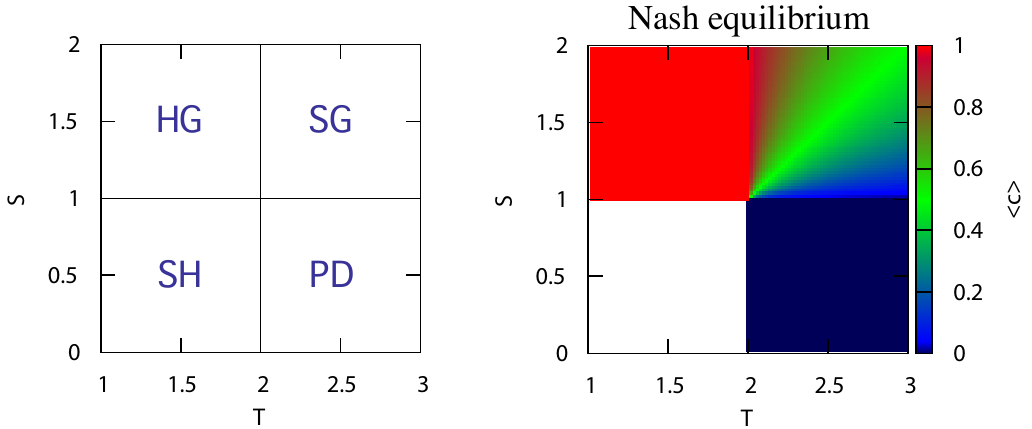}
\caption{Two dimensional ($S,T$) representation of the symmetric two-players two-actions games for $P=1$, $R=2$, $0<S<2$, and $1<T<3$. (\textbf{a}) Left panel shows the location of the four types of games. (\textbf{b}) Right panel shows (color code shown at right) the average level of cooperation in the corresponding Nash equilibrium.}
\end{figure}

\subsubsection{Cognitive hierarchy theory}

According to the cognitive hierarchy theory, each agent $i$ ($i=1,2,\ldots,N)$ is characterized by her cognitive level $l_i$ ($l_i = 0,1,2\ldots$) and her assumed distribution of other players' levels. Level-$0$ players ($l_i=0$) choose their actions randomly, which means that a level-$0$ player should cooperate with probability $p_c=1/2$, regardless of the values of the payoff matrix. 
A level-$1$ player ($l_i=1$)
assumes that the other players will act non-strategically (\textit{i.e.}, as level-$0$ players).
In the same way, a level-$h$ player ($h>1$) assumes a heterogeneous population consisted of 
players of lower levels $0,1,2,\ldots, h-1$. A strategic agent $i$ ($l_i>0$) assumes that
the cognitive levels of her $N-1$ opponents are distributed according to a given distribution (Camerer {\textit{et al.}} \cite{Camerer2004} considered this to be Poisson). In particular,
a level-$h$ player ($h>1$) assumes respective ratios $g_h(k)$ of
level-$k$ players, $k=0,1,\ldots,h-1$, with 

\begin{equation}
\sum_{k=0}^{h-1} g_h(k)=1\;\;.
\label{normalization}
\end{equation}

Then, each agent
chooses the action that would provide a higher payoff
if the cognitive levels of the rest of the agents were distributed according to her
assumption. The next subsection is devoted to the analysis of the actions taken by the agents in the four types of games under the assumptions of the cognitive hierarchy theory.

\subsection{Analysis}
\subsubsection{Harmony Game}

Provided $S>P,\, R>T$, the expected payoff is higher for cooperation regardless other players' actions. In consequence, all strategic players (level higher than zero) will choose cooperation. In the HG cooperation is the only strict best response to itself and to defection.

\subsubsection{Prisoner's Dilemma}

Given the payoff's ordering $T>R>P>S$, whatever the value of the cooperation level $c$ is, the expected payoff is higher for defection, and that is what a strategic player $i$ ($l_i>0$) should choose. In the PD game only the defective action is a strict best response to itself and to cooperation.

\subsubsection{Stag Hunt}

A player of level $1$ assumes a population consisting of $N-1$ opponents of level-$0$, that is, she assumes a fraction of cooperators $c=1/2$. According to equation (\ref{S_threshold_Tc}), a level-$1$ strategist playing a SH should cooperate if and only if:

\begin{equation}
S>S_{th}(T;c=1/2)=\frac{1+(1/2)(T-3)}{1-(1/2)}=T-1\;\;.
\label{Level1SH}
\end{equation}

Now a level-2 player has to consider two situations:
\begin{itemize}

\item[(i)] For $S>T-1$, we have $S>S_{th}(T;1/2)$ and level-$1$ players will cooperate. Thus, the average cooperation $c$ assumed by a level-$2$ player will be $c=g_2(0)/2+g_2(1)=g_2(0)/2+g_2(0)=1-g_2(0)/2$. Provided $g_2(0)<g_1(0)=1$, {\textit{i.e.}} level-$2$ players assume at least one level-$1$ player,
we have $c>1/2$, and therefore (using that, for $T<2$, $S_{th}$ is a decreasing function of $c$) $S_{th}(T;c)<S_{th}(T;1/2)$, which implies that a player of level $2$ playing a SH will choose to cooperate if $S>T-1$.

\item[(ii)] For $S<T-1$, we have $S<S_{th}(T;1/2)$ and level-$1$ players will defect. The assumed cooperation level $c$ is 
$c=g_2(0)/2$. Provided $g_2(0)<g_1(0)=1$, we have $c<1/2$, hence (as $T<2$) $S_{th}(T;c)>S_{th}(T;1/2)$, and thus a player of level
$2$ playing a SH will chose to defect if $S<T-1$.

\end{itemize}

Consequently, a level-2 player takes the same action as a level-1 player does: to cooperate if and only if $S>T-1$. Let us assume that level-$k$ players ($k=1, 2, \dots, h-1$) cooperate if and only if $S>T-1$. Then, a level-$h$ player will assume
 
\begin{equation}
c=\frac{g_h(0)}{2} +\sum_{k=1}^{h-1} g_h(k)=1 - \frac{g_h(0)}{2} > \frac{1}{2}\;\;,
\label{SH-level-h}
\end{equation}

\noindent so that she cooperates if and only if $S>T-1$, and thus the induction argument allows to conclude that all strategic players cooperate if and only if $S>T-1$ in the SH game.

Summarizing, the line $S=T-1$ divides the quadrant SH into two octants: In the upper octant ($S>T-1$), all players of level higher than zero cooperate, while in the lower one ($S<T-1$) such players defect. This result is general, for any kind of normalized distributions $g_l(k)$ ($k=0, \dots, l-1$; and $l\geq1$) assumed by the agents, and was already pointed out in \cite{Camerer2004}.

\subsubsection{Snowdrift Game}

A player of level-$1$ considers that the rest of the players play at random, so that she assumes $c=1/2$. In consequence, a level-$1$ strategist playing a SG should cooperate if and only if:

\begin{equation}
S>S_{th}(T;c=1/2)=\frac{1+(1/2)(T-3)}{1-(1/2)}=T-1\;\;.
\label{Level1SG}
\end{equation}

Note that this condition coincides with the cooperation condition (\ref{Level1SH}) for level-$1$ players playing a SH game. However, things are different for higher level players in the SG, as we now will see. From a technical point of view, the reason is that for the SG, where $T>2$, $S_{th}$ is an increasing function of $c$, reflecting a well-known feature of the Hawk-Dove formulation of the SG, namely that in a population of hawks (doves) it is advantageous to play dove (resp. hawk).

Again, a level-2 player has to consider two situations:
\begin{itemize}

\item[(i)] For $S>T-1$, we have $S>S_{th}(T;1/2)$ and level-$1$ players cooperate. Thus, the average cooperation $c$ assumed by a level-$2$ player will be $c=g_2(0)/2+g_2(1)=g_2(0)/2+g_2(0)=1-g_2(0)/2$. Provided $g_2(0) < g_1(0)=1$, {\textit{i.e.}} level-$2$ players assume at least one level-$1$ player,
we have $c > 1/2$, and therefore $S_{th}(T;c) > S_{th}(T;1/2)$, which implies that a player of level $2$ playing a SG will choose to cooperate if $ S > S_{th}(T;c)$, while she will choose to defect if $T-1 < S < S_{th}(T;c)$, with $c=1-g_2(0)/2$.

\item[(ii)] For $S<T-1$, level-$1$ players defect. Then, the assumed cooperation level $c$ is $c=g_2(0)/2$. Provided $g_2(0)<g_1(0)=1$, we have $c<1/2$, hence (as $T>2$) $S_{th}(T;c)<S_{th}(T;1/2)$. Thus a player of level $2$ will choose to cooperate if $S_{th}(T;c)<S<T-1$, while she will choose to defect if $S < S_{th}(T;c)$, with $c= g_2(0)/2$.

\end{itemize}

Consequently, regarding the action a level-2 player takes, there are four sectors in the SG quadrant ($T \in [2,3],\; S \in [1,2]$):

\begin{itemize}
\item[(a)] $1 < S < S_{th}(T;g_2(0)/2)$, defection.
\item[(b)] $S_{th}(T;g_2(0)/2)< S <T-1$, cooperation.
\item[(c)] $T-1 < S < S_{th}(T;1-g_2(0)/2)$, defection.
\item[(d)] $ S_{th}(T;1-g_2(0)/2) < S < 2$, cooperation.

\end{itemize}

Note that two of the borderlines separating these regions are dependent on the distribution assumed by the level-2 player, {\textit{i.e.}} these regions are non-universal.

At this point, one realizes that regarding the action a level-$l$ takes, there may appear more and more regions in the SG quadrant, depending on the specific assumption on the distributions $g_h(k)$ ($k=0, \dots, h-1$; and $l>h\geq1$). As an illustrative example, see Appendix A for the possibilities that arise for the actions taken by a level-3 player.

Despite this non-universality and increasing complexity with cognitive levels that characterize the actions taken by players of the SG, we show in the next sub(sub)section two general symmetries that universally hold, under the assumptions of the cognitive hierarchy theory.

\subsubsection{Symmetries in the Snowdrift Game}

As before, to simplify notation we will assume the values $P=1$ and $R=2$, though the arguments below remains valid for other values compatible with SG.

Given a particular SG game, corresponding to a pair of values $(T,S)$, with $T>2$ and $S>1$, we will say that it is a game of class $m$ whenever

\begin{equation}
m= \frac{S-1}{T-2}\;\;.
\label{game_m}
\end{equation}

\noindent In other words, $m$ is simply the slope of the straight line connecting the points $(T=2, S=1)$ and $(T, S)$.

The first statement that we will prove is the following:
\begin{itemize}
\item[\bf{S1}] Any two SG games of the same class $m$ are equivalent, in the sense that any player takes the same action in both games.
\end{itemize}

To prove this statement, note that Eq. (\ref{S_threshold_Tc}) can be rewritten as

\begin{equation}
S_{th}(T;c)=\frac{c}{1-c}T + \left(1-\frac{2c}{1-c}\right)\;\;,
\label{S_threshold_Tc2}
\end{equation}

\noindent so that a rational player playing a game of class $m$ cooperates if $m> c/(1-c)$, and defects if $m< c/(1-c)$. Here $c$ is the value of the average cooperation in the population estimated by the rational player under the assumption of a particular distribution of cognitive levels.

Now, the value of $c$ that a level-1 player estimates is $c=1/2$, irrespective of any consideration, so the action she takes is the same for all games in the same class. Consequently, the estimation of $c$ by a level-2 player is the same in all games of the same class, so that she takes the same action in all of them, and so on for all cognitive levels, which ends the proof of statement {\bf{S1}}.

To avoid possible misunderstandings, let us emphasize that the payoffs received by a player in two equivalent games can be very different. The notion of equivalence between games means here equality of the actions taken by an agent in both games, but it doesn't mean equality of payoffs received.

In what follows, a game $m$ is a game of class $m$. A second symmetry is the following:
\begin{itemize}
\item[\bf{S2}] The action that a player takes in the game $m$ is the opposite to the action she takes in the game $m^{-1}$.
\end{itemize}

Level -1 players satisfy trivially the statement {\bf{S2}}, for if $m>1$, then $m^{-1}<1$. Now, let us assume that for levels $1, \dots , l-1$ the statement holds. Let us call $C_l$ the subset of these levels whose actions in the game $m$ are cooperation, and $D_l$ its complementary. Then, level-$l$ players estimate

\begin{equation}
c= \frac{g_l(0)}{2} + \sum_{i \in C_l} g_l(i) \;\;,
\label{c_estimate_m}
\end{equation}

\noindent for the game $m$, while they estimate
 
\begin{equation}
c'= \frac{g_l(0)}{2} + \sum_{i \in D_l} g_l(i) = 1-c \;\;
\label{c_estimate_minverse}
\end{equation}

\noindent for the game $m^{-1}$, where the last equality follows from the normalization condition on the distribution of cognitive levels. Consequently, level-$l$ players satisfy statement {\bf{S2}}, for if $m>c/(1-c)$, then $m^{-1}<c'/(1-c')$. Thus the statement {\bf{S2}} is proved by the induction argument.

\subsection{Dynamics}

In this subsection we introduce a very simple dynamics for the temporal evolution of the distribution that each agent assumes on the cognitive levels of the population, and show results for this dynamics. The assumption is that the only information available to each agent $i$ at a given instant of time $t>1$ is her current payoff, $\Pi_i^t$, and her previous payoff, $\Pi_i^{t-1}$. Before the presentation of the dynamics, we briefly discuss the types of distribution of cognitive levels considered in the simulations performed.

\subsubsection{Distributions of cognitive levels}

The first type of distribution that we have considered (below referred to as scenario A) is the "normalized" (truncated) Poisson distribution employed in reference \cite{Camerer2004}, defined as follows. A Poisson distribution is described by a single parameter $\tau$, which is the mean and the variance:

\begin{equation}
f_{\tau}(n) = \frac{\tau^n e^{-\tau}}{n!} \;.
\label{Poisson}
\end{equation}

\noindent A strategic agent $i$ whose cognitive level is $l_i$ ($>0$) assumes a value of $\tau=\tau_i$, and that the cognitive levels $l_j$ ($=0, \dots, l_i-1$) of her opponents are distributed according to 

\begin{equation}
g^A_{l_i,\tau_i} (l_j)= \frac{f_{\tau_{i}}(l_j)}{C_i}\;\; ,
 \label{Poisson_truncated}
\end{equation}

\noindent where $f_{\tau}$ is the Poisson distribution (\ref{Poisson}), and $C_i$ is an appropriate normalization constant, {\em{i.e.}}

\begin{equation}
C_i= \sum_{k=0}^{l_i-1}f_{\tau_{i}}(k)\;\;\; . 
\label{normalization_constant}
\end{equation}

Writing equation (\ref{Poisson_truncated}) explicitly one has:

\begin{equation}
g^A_{l_i,\tau_i} (l_j) =\frac{\tau_i^{l_j}}{l_j!\sum_{k=0}^{l_i-1}\frac{\tau_i^k}{k!}}\;\;\; . 
\label{explicit_Poisson}
\end{equation}

A second type of cognitive levels distribution (scenario B) uses, instead of a Poisson distribution, the following exponential law:

\begin{equation}
f(n)=\frac{1}{2^{n+1}} \;\;. 
\label{exponential}
\end{equation}

\noindent Now, a strategic agent $i$ whose cognitive level is $l_i$ ($>0$) assumes that the cognitive levels $l_j$ ($=0, \dots, l_i-1$) of her opponents are distributed according to 

\begin{equation}
g^B_{l_i} (l_j)=\frac {f(l_i-l_j-1))}{Z_i}\;, \;\;  Z_i=\sum_{k=0}^{l_i-1} f(l_i-k-1) \;\;, 
\label{exponential_truncated}
\end{equation}

\noindent that is, explicitly:

\begin{equation}
g^B_{l_i} (l_j)=\frac {2^{l_j-l_i}}{\sum_{k=1}^{l_i}2^{-k}} \;\;. 
\label{explicit_exponential}
\end{equation}

The third type of distribution (scenario C) that we consider here is just a normalized uniform distribution:

\begin{equation}
g^C_{l_i}(l_j) = \frac{1}{l_i} \;\;
\label{uniform}
\end{equation}

\subsubsection{Dynamics algorithm}

One must first specify the initial condition ($t=0$) for the dynamics. In the simulations that we show below our choice is a population with cognitive levels $l_i$ ($0\leq l_i\leq l_{max}$) distributed according to a truncated Poisson distribution $g^A_{l_{max},\tau} (l_i)$ given by equation (\ref{explicit_Poisson}) where $\tau=1.5$ and $l_{max}=20$. For the cases in which the distribution of cognitive levels assumed by the agents is also "truncated Poisson", the initial rate parameter $\tau_i$ of an agent $i$ is taken to be $\tau_i=0$ if $l_i\leq 1$, and $\tau_i=\frac{l_i-1}{2}$ otherwise. Then the agents play simultaneously a one-shot game where the action taken by each strategic agent is the best response for her assumed distribution (either $g^A_{l_i,\tau_i} (l_j)$, or $g^B_{l_i} (l_j)$, or $g^C_{l_i}(l_j)$) for the cognitive levels of her opponents, each one receiving an initial payoff $\Pi_i(0)$.

Thereafter, the dynamics proceeds according to the following rules: at each time step $t>0$ 

\begin{itemize}
\item[Step 1] The agents play simultaneously with the action which is the best response according to their current beliefs (random for level-0 players), each one receiving a payoff $\Pi_i(t)$.
\item[Step 2] Each agent $i$ compares her current and previous payoff. If $\Pi_i(t) \geq \Pi_i(t-1)$, the agent $i$ keeps her current belief on the population distribution, while if $\Pi_i(t) < \Pi_i(t-1)$, the agent makes an attempt to change her belief.
\end{itemize}

The attempt to change the currently assumed distribution, for the cases in which this is $g^B_{l_i}$ or $g^C_{l_i}$ (say scenarios B or C) consists of two mutually exclusive possible events:

\begin{itemize}
\item With probability $u$ agent $i$ varies her level $l_i$ according to $l_i(t+1)=l_i(t)\pm 1$, that is, in an equiprobable way she increases or decreases its level $l_i$ at a point. 
\item Otherwise (\textit{i.e.}, with probability $1-u$), she keeps her cognitive level.
\end{itemize}

For the cases in which the agents assume a truncated Poisson distribution, $g^A_{l_i,\tau_i}$ (scenario A), the trial to change current beliefs consists of three mutually exclusive possible events:

\begin{itemize}
\item With probability $u$ agent $i$ varies her level $l_i$ according to $l_i(t+1)=l_i(t)\pm 1$, that is, in an equiprobable way she increases or decreases its level $l_i$ at a point. 
\item With probability $v$ (where $u+v\leq 1$) agent $i$ varies
her assumed rate parameter $\tau_i$ according to
$\tau_i(t+1)=\tau_i(t)+\epsilon$, where $\epsilon \in [-\delta,\delta]$, preserving $\tau_i \geq 0$.
\item Otherwise (\textit{i.e.}, with probability $1-u-v$), nothing changes.
\end{itemize}

Let us note that the presence of non-strategic (level-0) agents in the initial population is, within this dynamics, a necessary condition for a proper time evolution. A non-strategic agent chooses her action at random, and thus with probability $1/2$ her action at $t=1$ is different from that at $t=0$, then making possible that $\Pi_i(1) < \Pi_i(0)$ for some $i$. 

\subsubsection{Simulations results}

\begin{figure}[h]
\centering
\includegraphics[width=15cm]{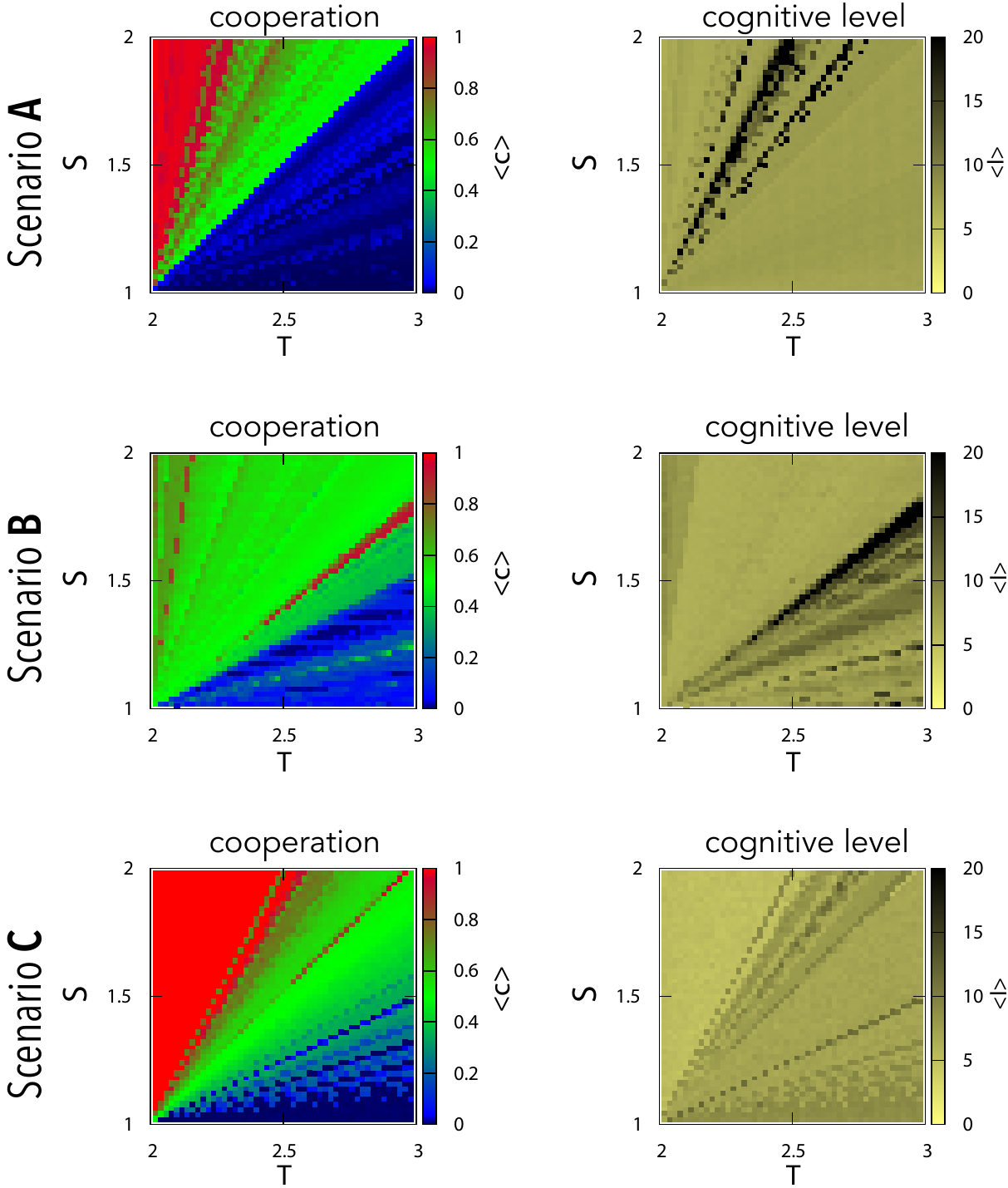}

\caption{Simulation of the dynamics of the SG game. The figure shows in color code representation the averaged values of cooperation (left panels) and cognitive level (right panels) in the stationary state for the initial conditions specified in main text. (\textbf{a}) Upper panels show the results for the scenario A, where agents assume a truncated Poisson distribution of cognitive levels in the population. (\textbf{b}) Mid panels show the results for the scenario B, and (\textbf{c}) lower panels show the results for an assumed homogeneous distribution (scenario C). Each point shows the result averaged over 100 Monte Carlo simulations. Parameter values are $N=10^3$, $u=0.45$, and (for the scenario A) $\delta=1$, and $v=0.45$.}
\label{f2}
\end{figure}

We have performed Monte Carlo simulations of the dynamics defined above for the cognitive hierarchy theory of the SG game, for the three scenarios A, B, and C that correspond, respectively, to the agents' assumption for the cognitive levels distribution given by $g^A$ (equation (\ref{explicit_Poisson})), $g^B$ (equation (\ref{explicit_exponential})), and $g^C$ (equation (\ref{uniform})). In all the cases, the initial conditions were as described in previous subsection, {\textit{i.e.}} cognitive levels were initially distributed in the population according to a truncated Poisson distribution with $\tau=1.5$ and $l_{max}=20$. The population size is $N=10^3$, the probability of changing the current cognitive level (provided payoff decreases) is $u=0.45$, and, for the scenario A, 
$\delta=1$, and $v=0.45$.

In figure \ref{f2} we show, for the three scenarios, the averaged value over one hundred simulations (for each ($T,S$) point) of the fraction of cooperators and the average cooperative level in the stationary state reached by the dynamics in the whole SG quadrant ($T \in [2,3],\; S \in [1,2]$).

From the inspection of figure \ref{f2}, a visible result is that the symmetry {\bf{S1}} (equivalence of games in the same class $m$) is near preserved by the dynamics. The result is indeed remarkable, in the extent that the preservation of this symmetry requires that some specific conditions hold, so that the symmetry conservation is non-generic. This is discussed in detail in Appendix B, where those specific conditions are derived. On the contrary, figure \ref{f2} clearly shows the breaking of the symmetry {\bf{S2}} (mirror anti-symmetry respect to the main diagonal of the SG quadrant), in full agreement with the analysis of this symmetry in Appendix B, which shows that no specific conditions are needed for the breaking of this symmetry.

Regarding the cooperation level reached for the different scenarios, the differences are also remarkable. In scenario A near full defection largely dominates below the main diagonal, with a sharp change above it to cooperation values larger than $1/2$ which show an overall tendency to increase with the value $m$ of the equivalence class slope, and near full cooperation as $m \rightarrow \infty$. In contrast, in scenario B, neither full defection nor full cooperation are almost present (except for a few tiny sectors), with intermediate values of cooperation being largely predominant. For scenario C, a state of full cooperation ($c=1$) is reached for $m>2$, while almost fully defective states are only seen for $m<1/2$, becoming dominant only as $m \rightarrow 0$.

\begin{figure}[h]
\centering
\includegraphics[width=9cm,angle=-90]{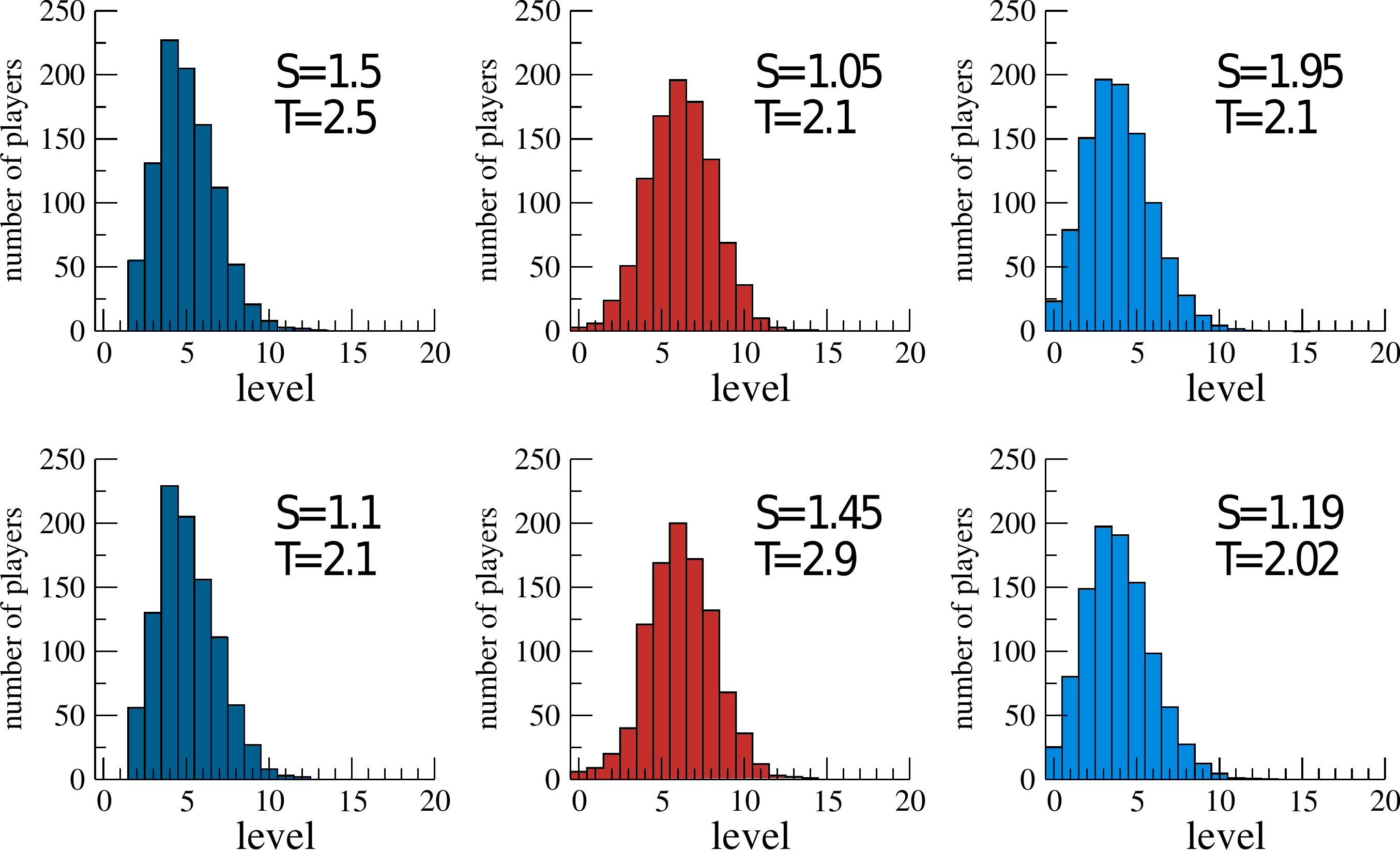}
\caption{Histograms of the cognitive levels of the population in the stationary state of the dynamics of the SG game in the scenario A. Panels in the same vertical correspond to games in the same equivalence class $m$. The values of the game parameters $T$ and $S$ are shown inside the corresponding panel.}
\label{f3}
\end{figure}   

In figure \ref{f3} we represent the histograms, for a few selected points of the SG quadrant ($T \in [2,3],\; S \in [1,2]$), of the cognitive levels in the stationary state for the scenario A. Each point ($T,S$) of a lower panel belongs to the equivalence class of the correlative upper panel, to show the near preservation of the symmetry {\bf{S1}}.

\begin{figure}[h]
\centering
\includegraphics[width=9cm,angle=-90]{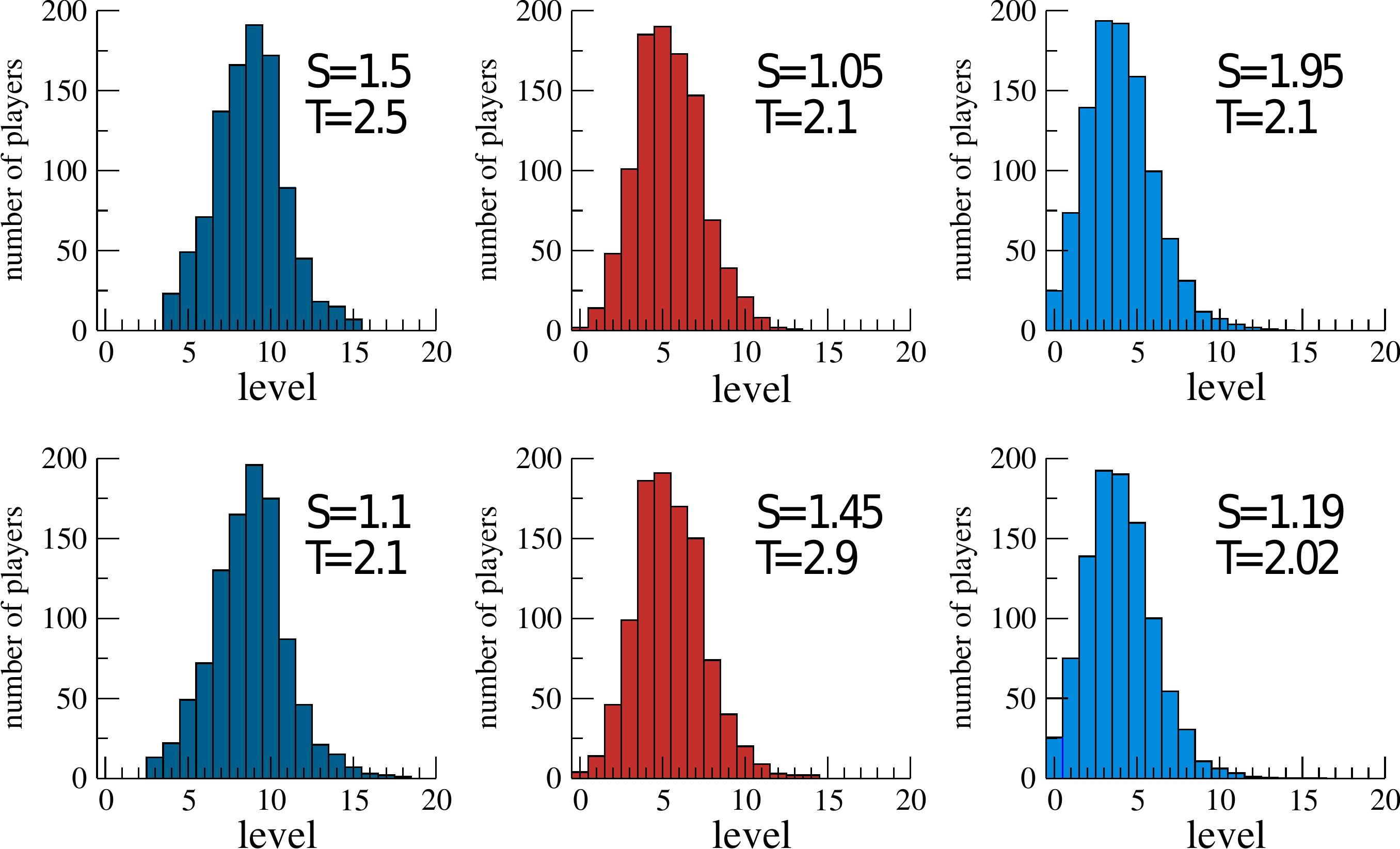}
\caption{Histograms of the cognitive levels of the population in the stationary state of the dynamics of the SG game in the scenario B. Panels in the same vertical correspond to games in the same equivalence class $m$. The values of the game parameters $T$ and $S$ are shown inside the corresponding panel..}
\label{f4}
\end{figure}

\begin{figure}[h]
\centering
\includegraphics[width=9cm,angle=-90]{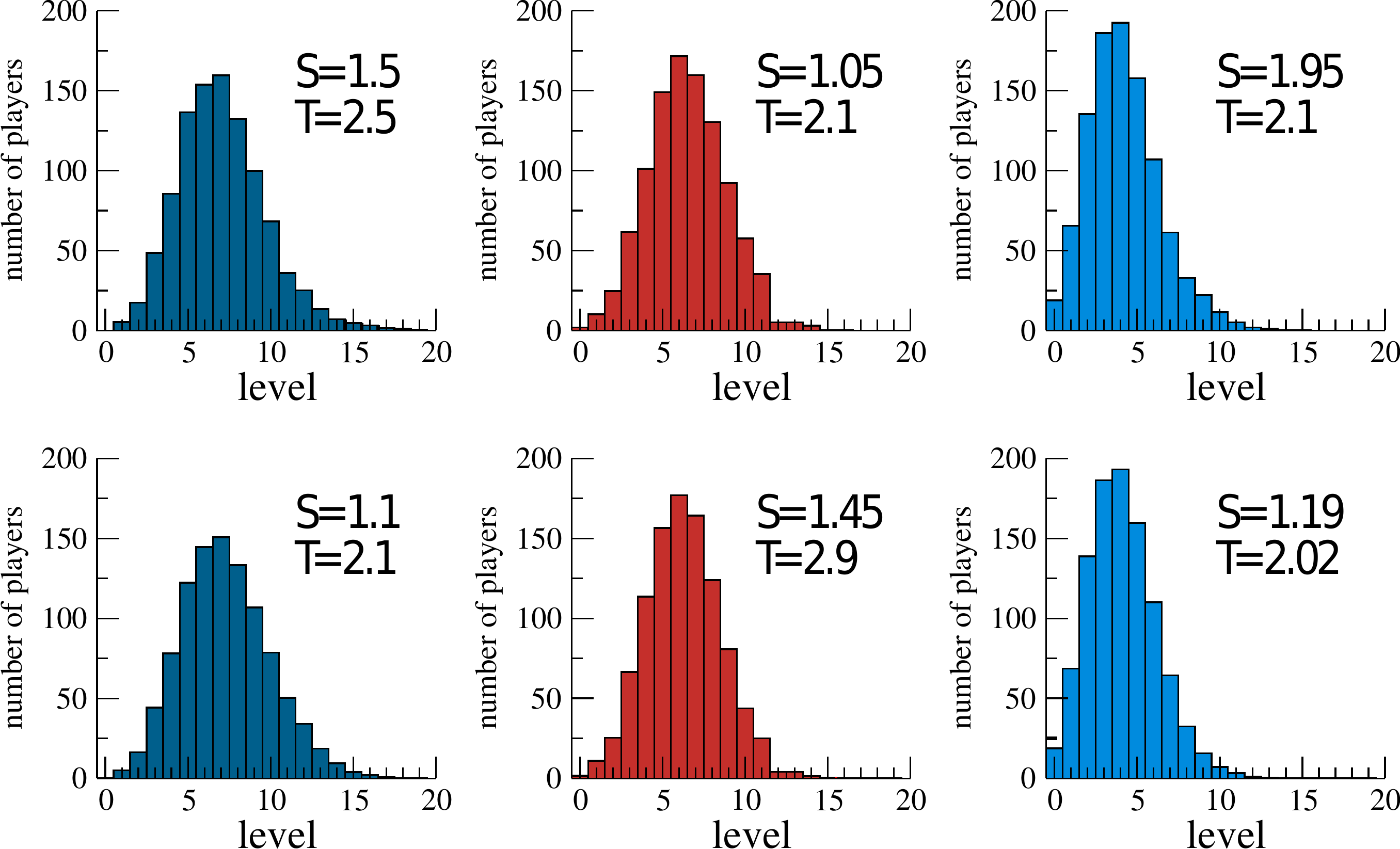}
\caption{Histograms of the cognitive levels of the population in the stationary state of the dynamics of the SG game in the scenario C. Panels in the same vertical correspond to games in the same equivalence class $m$. The values of the game parameters $T$ and $S$ are shown inside the corresponding panel.}
\label{f5}
\end{figure}

Figures \ref{f4} and \ref{f5} are as figure \ref{f3}, but for the scenarios B and C, respectively. The differences between the three figures are merely of a quantitative nature, as indeed they exhibit the same main qualitative features. This observation points out to the conclusion that the qualitative aspects of the distribution of cognitive levels in the stationary state of this dynamics is, to a large extent, rather insensitive respect to the agents' beliefs. This should undoubtably be ascribed to the very scarce information (own current and previous payoff) available to agents in this dynamics. On the other hand, this insensitivity is in contrast with the large differences in the average cooperation reached for the three scenarios, as observed in figure \ref{f2}. However this is in no way contradictory: Even for an identical distribution of agents' cognitive levels in the population, as far as the agents conform their actions to their beliefs (and not to the real distribution, which they ignore), different scenarios ({\textit{i.e.}} different beliefs) produce different cooperation patterns.

The Poisson-like aspect of the histograms in figures \ref{f3}, \ref{f4} and \ref{f5} may suggest that, given that our initial condition for the cognitive levels distribution is truncated Poisson, the dynamics preserves the type of initial distribution, with perhaps some shift and small deformation. However this conjecture is invalidated by simulations (not shown) performed with other non-Poisson initial distributions of cognitive levels: even for an initial uniform distribution, unimodal histograms are quickly observed to emerge from the dynamics.

\begin{figure}[h]
\centering
\includegraphics[width=15cm]{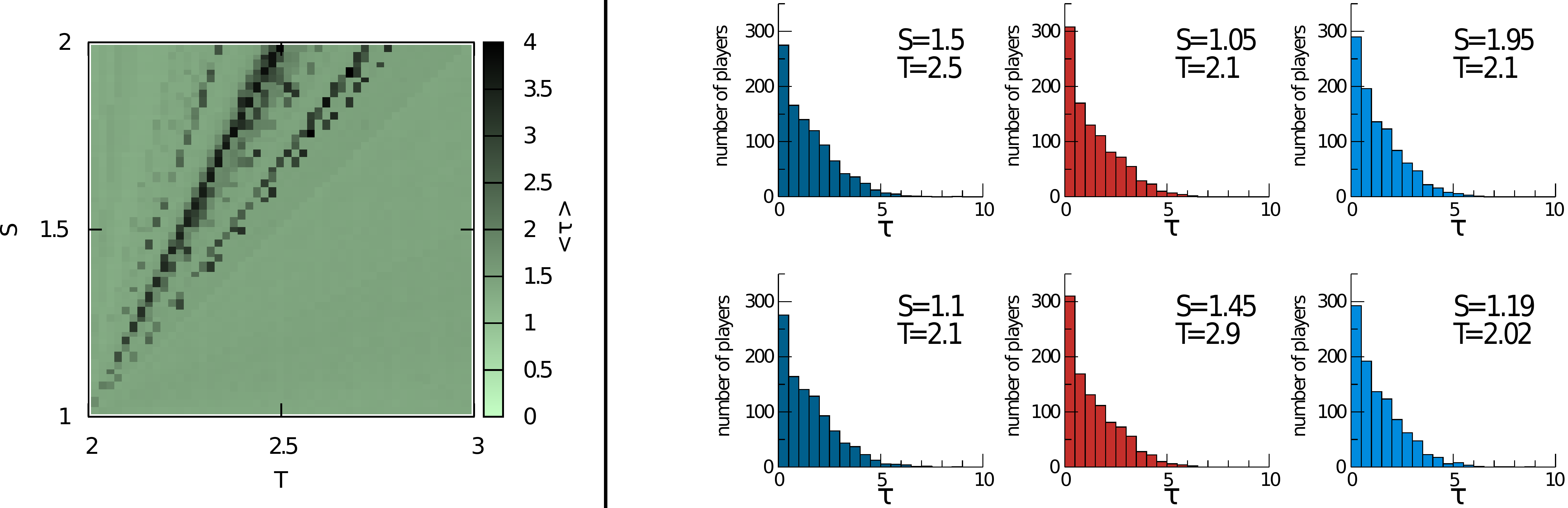}
\caption{Dynamics of the SG game in the scenario A. (\textbf{a}) Leftmost panel shows in color code the average $\langle \tau \rangle$ of the Poisson parameter of the population in the stationary state. (\textbf{b}) Panels in the right show the histograms of $\tau$ for a few selected points, with figures in the same vertical corresponding to games in the same equivalence class $m$. The values of the game parameters $T$ and $S$ are shown inside the corresponding panel.}
\label{f6}
\end{figure}

In scenario A, where the agents believe that the cognitive levels in the population are distributed following a truncated Poisson distribution, not only the cognitive levels evolve, but also the Poisson parameters $\tau_i$ do. So, how do they evolve? Figure \ref{f6} show, in the leftmost panel, the average $\langle \tau \rangle$ of the Poisson parameter of the population in the stationary state. A clear correlation of this quantity with the averaged cooperation level shown in the upper rightmost panel of figure \ref{f2} is observed. In the right part of figure \ref{f6} we show, for the same points ($T,S$) used in previous figures, the histograms of the Poisson parameter values in the stationary states.

\section{Conclusions}

We have analyzed here the cognitive hierarchy theory for agents playing two-person two-action games in a well-mixed population. While for the HG, PD and SH games the results are straightforward and universal, {\textit{i.e.}} independent of the specific distribution of cognitive levels assumed by the agents, for the SG game the analysis show an increasing complexity with cognitive levels, with results that are non-universal, in the sense that the actions taken by the high cognitive level agents depends on the specificities of the assumed distribution. Despite this non-universality, we find two exact symmetries: For a given assumed distribution of cognitive levels, agents of a fixed cognitive level take the same action (symmetry {\bf{S1}}) in all the games ($T,S$) sharing the value of $m=(S-1)/(T-2)$, while they take the opposite action (symmetry {\bf{S2}}) in all the games ($T',S'$) with $(S'-1)/(T'-2) = m^{-1}$. 

We introduce a stochastic dynamics where agents can update their current beliefs on the distribution of cognitive levels in the population, with no available information other than their current and previous payoffs. The simulations of the SG game for this dynamics converge to stationary states of the population characterized by an average fraction of cooperators which depends largely on the agents beliefs, but where in contrast, the distribution of cognitive levels reached is rather insensitive to their beliefs.

We provide arguments showing that for synchronous updating, the previous dynamics breaks forcefully the symmetry {\bf{S2}}, while the breaking of the symmetry {\bf{S1}} requires some specific conditions, so that though its preservation is non-generic, nonetheless it is not forbidden. Our simulations for different scenarios show the breaking of the symmetry {\bf{S2}}, and an apparent conservation of the symmetry {\bf{S1}}.

\acknowledgments{\textbf{Acknowledgments:} We acknowledge financial support from the European Commission through FET IP project MULTIPLEX (Grant No. 317532), from the Spanish MINECO under project FIS2014-55867-P, from the Departamento de Industria e Innovaci\'on del Gobierno de Arag\'on y Fondo Social Europeo (FENOL group E-19),}

\appendix 

\section*{\noindent Appendix A}\vspace{6pt} 

As an illustration of the increasing complexity, and non-universality, of the analysis of the actions taken by players of high cognitive levels in the SG game, we will analyze here the case of level-3 players. Regarding the assumed distribution of cognitive levels by a level-3 player, the only assumption we make is that $g_3(0),g_3(1),g_3(2)<1$, that is,  players of level $3$ assume that not all the other players belong to the same level.

Let us first remind here the results of 2.2.4 concerning the actions of lower level players:
\begin{itemize}
\item[1.] Level-1 players cooperate if and only if $S>T-1$.\vspace{2pt}

\item[2.] Level-2 players take the following actions, depending on the point $(T,S)$ representing the particular SG game:
\begin{itemize}
\item[(a)] $1 < S < S_{th}(T;g_2(0)/2)$, defection.
\item[(b)] $S_{th}(T;g_2(0)/2)< S <T-1$, cooperation.
\item[(c)] $T-1 < S < S_{th}(T;1-g_2(0)/2)$, defection.
\item[(d)] $ S_{th}(T;1-g_2(0)/2) < S < 2$, cooperation.
\end{itemize}

\end{itemize}

Due to the existence of the symmetry {\bf{S2}} (see above in 2.2.5), we can restrict consideration to regions (a) and (b), that is, the octant $1<S<T-1$.

In region (a) level-1 and level-2 players defect. Hence, level-3 players assume a cooperation level $c_a=g_3(0)/2$. Therefore,  the condition for level-$3$ players to cooperate in this region becomes $S>S_{th}(T;g_3(0)/2)$, otherwise (\textit{i.e.},  if $S<S_{th}(T;g_3(0)/2)$) level-3 players defect. 
Now, if it is the case that $g_3(0)<g_2(0)$, then $S_{th}(T;g_3(0)/2)<S_{th}(T;g_2(0)/2)$ and therefore a level-3 player cooperates in the subregion $S_{th}(T;g_3(0)/2)<S<S_{th}(T;g_2(0)/2)$, while she defects in the complementary subregion $1<S<S_{th}(T;g_3(0)/2)$. 
On the other hand, if $g_3(0)>g_2(0)$ level-3 players always defect in region (a). Let us refer to the condition $g_3(0)<g_2(0)$ as {\bf{a.1}} and to its contrary as {\bf{a.2}}.

In region (b) level-1 players defect and level-2 players cooperate, so that the cooperation assumed by level-3 players is $c_b=g_3(0)/2+g_3(2)=1-g_3(0)/2-g_3(1)$. Consequently, a player of level-$3$ cooperates if and only if $S>S_{th}(T;1-g_3(0)/2-g_3(1))$. Now, if it is the case that $c_b=1-g_3(0)/2-g_3(1)<g_2(0)/2$ (condition {\bf{b.1}}), a level-$3$ player always cooperate in region (b); while if $c_b=1-g_3(0)/2-g_3(1)>1/2$ (condition {\bf{b.2}}) she always defect in region (b); finally if none of these conditions hold, namely if $g_2(0)/2 < c_b<1/2$ (condition {\bf{b.3}}), a level-3 player defects in the subregion $S_{th}(T;g_2(0)/2)< S <S_{th}(T;c_b)$, but she cooperates in the complementary subregion  $S_{th}(T;c_b)< S <T-1$.

\begin{table}[h]
\caption{The five possible scenarios for the actions taken by level-3 players of the SG game in the octant $1<S<T-1$. See main text for explanations on the notation.}
\small 
\centering
\begin{tabular}{ccc}
\hline
\textbf{Scenario}	& \textbf{Regions}	& \textbf{Action}\\
\hline
{\bf{a.1}} AND {\bf{b.1}}	& $1<S<S_{th}(T;g_3(0)/2)$	& D \\
		& $S_{th}(T;g_3(0)/2)<S<T-1$			& C\\
\hline
{\bf{a.1}} AND {\bf{b.2}}	& $1<S<S_{th}(T;g_3(0)/2)$	& D \\
		& $S_{th}(T;g_3(0)/2)<S<S_{th}(T;g_2(0)/2)$			& C\\	
		& $S_{th}(T;g_2(0)/2)<S<T-1$			& D\\	
\hline
{\bf{a.1}} AND {\bf{b.3}}	& $1<S<S_{th}(T;g_3(0)/2)$	& D \\
		& $S_{th}(T;g_3(0)/2)<S<S_{th}(T;g_2(0)/2)$			& C\\	
		& $S_{th}(T;g_2(0)/2)<S<S_{th}(T; c_b)$			& D\\	
		& $S_{th}(T;c_b)<S<T-1$			& C\\	
\hline
{\bf{a.2}} AND {\bf{b.2}}	& $1<S<T-1$	& D \\
\hline
{\bf{a.2}} AND {\bf{b.3}}	& $1<S<S_{th}(T;c_b)$	& D \\
		& $S_{th}(T;c_b)<S<T-1$			& C\\	
\hline
\end{tabular}
\end{table}

Summarizing the discussion, there can be {\textit{a priori}} up to six different scenarios for the actions taken by level-3 players in the octant $1<S<T-1$, that correspond to the six possibilities [{\bf{a.i}} AND {\bf{b.j}}], ({\bf{i}}$=1,2$, {\bf{j}}$=1,2,3$). But one of them, namely [{\bf{a.2}} AND {\bf{b.1}}], can never occur for it would imply $g_3(0) + g_3(1) >1$, thus violating the normalization condition on $g_3$. The other five scenarios are perfectly possible; indeed one can easily construct examples of particular distributions of cognitive levels for each one of them.

\section*{\noindent Appendix B}\vspace{6pt} 

The question addressed in this appendix is whether or not the dynamics introduced above preserves the symmetries {\bf{S1}} and  {\bf{S2}} of the cognitive hierarchy theory of the SG game. 

The preservation of the symmetry  {\bf{S1}} requires that the decision of every agent $i$ at any time $t$ of changing her beliefs be the same for all the games in the same class $m$ of equivalence, {\em{i.e.}}, that the sign of 

\begin{equation}
\Delta(S,m) =  \Pi_i(t+1; S,m) - \Pi_i(t; S,m)\;\;,
\label{Delta}
\end{equation}

\noindent for fixed $m$, is independent of $S$. 


Let us first consider an agent whose action at both times, $t$ and $t+1$, is cooperation, and let $c_0$ and $c_1$ be the fraction of cooperators at $t$ and $t+1$, respectively. Thus the payoff difference $\Delta$ is

\begin{equation}
\Delta(S,m) =  (c_1-c_0) (2-S)\;\;,
\label{case_CC}
\end{equation}

\noindent whose sign is then independent of $S$, provided we restrict consideration to $S<2$. Note that without this (somewhat arbitrary) restriction, symmetry {\bf{S1}} would already be broken for this simple case, whenever $\delta_c =c_1-c_0 \neq 0$. 

The analysis for the case of an agent whose action at both times, $t$ and $t+1$, is defection is also straightforward. The payoff difference $\Delta$ is

\begin{equation}
\Delta(S,m) =  (c_1-c_0) (T-1)= \delta_c (m^{-1} (S-1) +1)\;\;,
\label{case_DD}
\end{equation}

\noindent whose sign is then independent of $S$, for fixed $m$. In this case, symmetry  {\bf{S1}} is always preserved with no need of restriction on the $S$ (and $m$) values compatible with the SG game.

Let us now consider the case of an agent that cooperates at time $t$ but defects at time $t+1$. The payoff difference $\Delta$ is
now 

\begin{eqnarray}
\Delta(S,m) & = &  c_1(T-1) +1 - c_0(2-S) -S  \nonumber \\
 & = & \delta_c + (S-1)(c_0(1+m^{-1}) + m^{-1}\delta_c -1)\;\;.
\label{case_CD}
\end{eqnarray}

If $\delta_c =0$, then the sign of $\Delta$ is independent of $S$, for fixed $m$. Indeed, $\Delta$ is negative if and only if $c_0 < m/(m+1)$.

However, for $\delta_c \neq 0$ there is a change of sign in $\Delta(S,m)$, for fixed $m$, at a value of $S=S_c(c_0,\delta_c, m)$, given by

\begin{equation}
S_c(c_0,\delta_c,m) = 1 + \frac{\delta_c}{ 1-c_0(1+m^{-1})-m^{-1}\delta_c}\;\;,
\label{critical_CD}
\end{equation}

\noindent provided $S_c(c_0,\delta_c, m)>1$. If this is the case, the symmetry {\bf{S1}} is broken. In fact, it is easy to find particular values of $m$, $c_0$ and $c_1$ for which this condition holds, even with the (somewhat arbitrary) restriction to values of $S_c<2$.

Finally, for an agent that defects at time $t$ but cooperates at time $t+1$, the payoff difference $\Delta$ is

\begin{eqnarray}
\Delta(S,m) & = &  c_1(2-S) +S - c_0(T-1) -1  \nonumber \\
 & = & \delta_c + (S-1)(1-c_0(1+m^{-1}) -\delta_c)\;\;.
\label{case_DC}
\end{eqnarray}

As in the previous case, if $\delta_c =0$, then the sign of $\Delta$ is independent of $S$, for fixed $m$. In this case $\Delta$ is negative if and only if $c_0 > m/(m+1)$.

For $\delta \neq 0$ there is a change of sign in $\Delta(S,m)$, for fixed $m$, at a value of $S=S_c(c_0,\delta_c, m)$, given by

\begin{equation}
S_c(c_0,\delta_c,m) = 1 + \frac{\delta}{c_0(1+m^{-1})+\delta_c-1}\;\;,
\label{critical_DC}
\end{equation}

\noindent provided $S_c(c_0,\delta_c, m)>1$.

Summarizing, the dynamics introduced above doesn't preserves generically the symmetry {\bf{S1}} of the SG game. However, if the updating is asynchronous, where $\delta_c = c_1-c_0 = 0$, under the usual restriction of the values of the parameter $S<2$, the symmetry {\bf{S1}} is preserved. It should be emphasized, that when the updating is synchronous, the breaking of the symmetry requires certain conditions to hold for some agent at some time during the evolution, so that the observation of symmetry preservation is not forbidden "a priori".

To address the preservation of the symmetry {\bf{S2}} we will analyze now the updating of an agent in two SG games whose representative points in the diagram ($S,T$) are mirror images each other respect to the principal diagonal of the SG quadrant. If we denote by ($S, m$) one of the games, the other is ($S', m^{-1}$), with $S' = m^{-1}(S-1) +1$.

We assume that at a time instant $t$, the strategic configurations in both games are {\bf{S2}}-symmetric, so that the action taken by any agent in one of the games is the opposite she takes in the other, and the same occurs at time $t+1$. Then if $c_0$ and $c_1$ are, respectively, the fraction of cooperators at times $t$ and $t+1$ for the game ($S, m$), the values corresponding to the game ($S', m^{-1}$) are $c'_0= 1-c_0$ and $c'_1= 1-c_1$.

Let us first consider an agent that cooperates, at both times $t$ and $t+1$, in game ($S, m$), so that she defects at $t$ and $t+1$ in the mirror-symmetric game. The payoff differences are

\begin{eqnarray}
\Delta(S,m) & = &  (c_1 - c_0)(2-S)  \nonumber \\
\Delta(S', m^{-1}) & = & (c'_1-c'_0)(T'-1) = (c_0 - c_1)S\;\;.
\label{S2_case_CC}
\end{eqnarray}

\noindent Under the usual restriction, $S<2$, we see that ${\text{sign}}\; \Delta(S,m) = - {\text{sign}}\; \Delta(S', m^{-1})$, so that the updating decisions are opposite, and the symmetry {\bf{S2}} is broken, whenever $\delta_c =c_1-c_0 \neq 0$. Note that if $\delta_c=0$, both differences are zero and, in both games, the agent doesn't try updating. 

Let us now consider the case of an agent that cooperates at time $t$ but defects at time $t+1$ in game ($S, m$), so that she defects at $t$ and cooperates at $t+1$ in the mirror-symmetric game. The payoff differences are

\begin{eqnarray}
\Delta(S,m) & = &  \delta_c + (S-1)(c_0(1+m^{-1}) + m^{-1}\delta_c -1) \\
\Delta(S', m^{-1}) & = & \delta'_c + (S'-1)(1-c'_0(1+m) -\delta'_c)\;\;
\label{S2_case_CD}
\end{eqnarray}

\noindent where we called $\delta'_c = c'_1 - c'_0$, and used equation (\ref{case_CD}) for the first equation, and adapted equation (\ref{case_DC}) for the last one. To proceed further, one can use $\delta'_c = - \delta_c$, $c'_0 = 1-c_0$, $c'_1= 1-c_1$, and $S' = m^{-1}(S-1) +1$, to obtain

\begin{equation}
\Delta(S', m^{-1}) =\Delta(S,m) -\delta_c + c_0(1+m) - m\;\;.
\label{compare}
\end{equation}

Thus, if it is the case that $-\delta_c + c_0(1+m) - m >0$ the payoff differences have opposite sign for $\delta_c - c_0(1+m) + m < \Delta(S,m) <0$, while if $-\delta_c + c_0(1+m) - m <0$ the payoff differences have opposite sign for $0< \Delta(S,m) < \delta_c - c_0(1+m) + m$.

Let us first consider the case $\delta_c = 0$. If $c_0(m+1)-m >0$, then $\Delta(S, m) = (S-1)(c_0(1+m^{-1})-1) >0$ and thus the payoff differences have the same sign. While if $c_0(m+1)-m <0$, then $\Delta(S, m) = (S-1)(c_0(1+m^{-1})-1) <0$ and the payoff differences have also the same sign. Consequently, for $\delta_c=0$, the symmetry {\bf{S2}} is preserved.

For the case $\delta_c>0$, one can has

\begin{itemize}
\item If $-\delta_c + c_0(1+m) - m >0$, then $\delta_c +c_0(m+1)-m > 2\delta_c$, thus $\Delta(S,m) > (2T-3)\delta_c >0$, and the payoff differences have the same sign, and the symmetry {\bf{S2}} is preserved.
\item If $-\delta_c + c_0(1+m) - m <0$, the symmetry is preserved provided $c_0(m+1) - m < -\delta_c (T-1)/(T-2)$.
\end{itemize}

For the case $\delta_c<0$, one can has

\begin{itemize}
\item If $-\delta_c + c_0(1+m) - m <0$, then $\delta_c +c_0(m+1)-m < 2\delta_c$, thus $\Delta(S,m) < (2T-3)\delta_c <0$, and the payoff differences have the same sign, and the symmetry {\bf{S2}} is preserved.
\item If $-\delta_c + c_0(1+m) - m >0$, the symmetry is preserved provided $c_0(m+1) - m > -\delta_c (T-1)/(T-2)$.
\end{itemize}

Summarizing, the symmetry {\bf{S2}} is preserved for $\delta_c =0$, but it is always broken for $\delta_c \neq 0$.

\end{document}